# High-Speed Photonic Neuromorphic Computing Using Recurrent Optical Spectrum Slicing Neural Networks

K. Sozos[1], A. Bogris[1], P. Bienstman[2], G. Sarantoglou[3], S. Deligiannidis[1], C. Mesaritakis[3]

[1]Dept. of Informatics and Computer Engineering, University of West Attica, Aghiou Spiridonos, 12243, Egaleo, Greece

[2]Dept. of Information Technology, Ghent University-imec, Technologiepark Zwijnaarde 126, 9052 Gent, Belgium

[3]Dept. Information and Communication Systems Engineering, Engineering School, University of the Aegean, Palama 2, 83200, Samos, Greece

**Abstract:** Neuromorphic Computing implemented in photonic hardware is one of the most promising routes towards achieving machine learning processing at the picosecond scale, with minimum power consumption. In this work, we present a new concept for realizing photonic recurrent neural networks and reservoir computing architectures with the use of recurrent optical spectrum slicing. This is accomplished through simple optical filters placed in an loop, where each filter processes a specific spectral slice of the incoming optical signal. The synaptic weights in our scheme are equivalent to filters' central frequencies and bandwidths. This new method for implementing recurrent neural processing in the photonic domain, which we call Recurrent Optical Spectrum Slicing Neural Networks (ROSS-NNs), is numerically evaluated on a demanding, industry-relevant task such as high baud rate optical signal equalization (>100 Gbaud), exhibiting ground-breaking performance. The performance enhancement surpasses state-of-the-art digital processing techniques by doubling the reach while minimizing complexity and power consumption by a factor of 10 compared to state-of-the-art solutions. In this respect, ROSS-NNs can pave the way for the implementation of ultra-efficient photonic hardware accelerators tailored for processing high-bandwidth (> 100 GHz) optical signals in optical communication and high-speed imaging applications.

Recurrent Neural Networks (RNNs) are universal computational tools tailored to process time-dependent data [1]. State-of-the-Art RNN architectures, such as Long Short-Term Memory (LSTM), Bi-directional RNNs or Gated-Recurrent-Units (GRU) [2,3] remain notoriously difficult to train, requiring the optimization of a significant number of hyper-parameters. Furthermore, the practicality of RNNs becomes even more questionable, when multi-GHz data inference is required by demanding applications in the area of optical communications and imaging. Up to now, RNN's superiority over other nonlinear digital signal processing techniques has been proved only through offline signal processing. Unfortunately, their realization by field programmable gate arrays (FPGAs) or application specific integrated circuits (ASICs) constitutes a quite challenging task, especially if processing rates exceeding 50 GSa/sec are targeted [4]. Aiming to amend these drawbacks, Reservoir Computing (RC) has emerged as a neuromorphic paradigm that offers radical simplification of the cumbersome RNN training [5]. In detail, by splitting the recurrent network in a reservoir (hidden layer) with random and untrained weights and a readout layer, where all training is taking place in a linear manner, RC achieves significant complexity reduction, while retaining performance. Moreover, from a hardware perspective, the randomness of the reservoir layer does not translate to performance deterioration, but on the contrary provides robustness against fabrication imperfections. These unique features of RCs render them extremely hardware friendly for a significant number of implementations, exploiting diverse platforms ranging from spintronics [6], polaritons, CMOS electronics [7] to free-space optics [8,9] and integrated photonic based approaches [10]. Especially photonics technology constitutes a proliferating platform for such schemes, due to inherent advantages such as computational parallelism through signal multiplexing, low power consumption, high-bandwidth support and processing at the speed of light [10]. These merits are exploited to the maximum in applications where the information to be processed is already in the optical domain, therefore direct complex processing can be obtained, alleviating the need for power hungry opto-electronic and electro-optical conversions. On the other hand, while photonics is extremely suitable for implementing linear transformations using passive components [11,12], it fails to provide integrated and low-power non-linear nodes, which is a critical part of an RC/RNN architecture.

In a photonic RC context, most efforts have concentrated on the rich non-linear dynamics of semiconductor lasers subjected to feedback. These schemes, when combined with time-multiplexing, have proved their efficacy in addressing difficult problems like time-series prediction [13] and non-linear channel equalization or chromatic dispersion (CD) compensation in intensity modulation/direct detection (IM/DD) transmission systems [14]. This sub-category of photonic RC, called time-delayed RC in the literature, has minimum photonic hardware requirements, consisting of a single nonlinear physical node and multiple time-multiplexed virtual nodes. Nevertheless, it is not compatible with all-optical coherent processing and is not integration friendly due to the fact that the number of nodes is proportional to the length of external delay path. For the same reason, time-delayed systems may achieve real-time signal processing only up to 20 Gsa/s as, the smaller the symbol period, the lower the number of virtual nodes that can be exploited for processing, thus affecting processing power. To make matters worse, in time-delayed RCs, a high-speed pseudo-random generator is also needed so as to mask the incoming signal, thus evoking differentiation between the dynamics of the virtual nodes. This unavoidable requirement hinders all-optical implementations and increases the digital processing requirements. A different implementation strategy consists of RC or RNN with spatially distributed nodes that usually contain passive waveguides [15], semiconductor optical amplifiers [16] or micro-ring resonators (MRRs) [17,18]. In terms of a node's non-linearity, a limited number of solutions have been proposed, either demanding power-hungry active elements, power-demanding nonlinear phase shifters relying on Kerr effect or are based on the square law offered by the photodiodes at the output layer. The use of photonic components for realizing RC nodes offers practically unlimited processing speed without sacrificing coherent processing. This feature outweighs all spatial RC's restrictions when high-speed applications are considered. In this context, photonic RC has attracted attention in the optical communications field,
thanks to its ability to compensate transmission impairments such as CD-induced power fading and Kerr-related non-linearities [19].

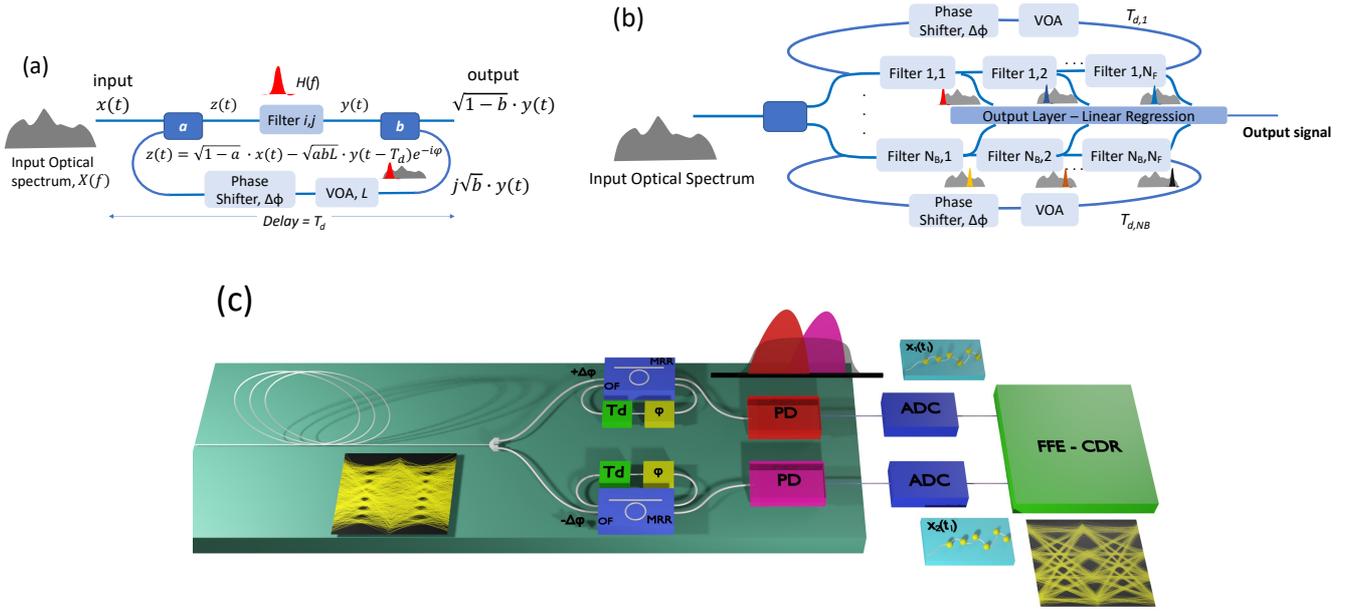

**Fig. 1. The architectural structure of ROSS-NN.** a. Configuration of a single ROSS-NN node, b. The architecture of ROSS-N consisting of $N_B$ filter banks with each bank consisting of $N_F$ filters in-a-loop serving as recurrent optical spectrum slicers. This architecture has been benchmarked in emulating the NARMA-10 sequence. c. The architecture of ROSS-N as a hardware neuromorphic processor for high-speed optical communications signals suffering from chromatic dispersion, bandwidth limitations of the transceiver and nonlinear effects. Two photodiodes receive two optical spectral slices of the signal recurrently processed by ROSS-NN. With the use of a feed-forward equalizer of fairly low complexity, significant restoration of signal fidelity is achieved according to thorough numerical simulations.

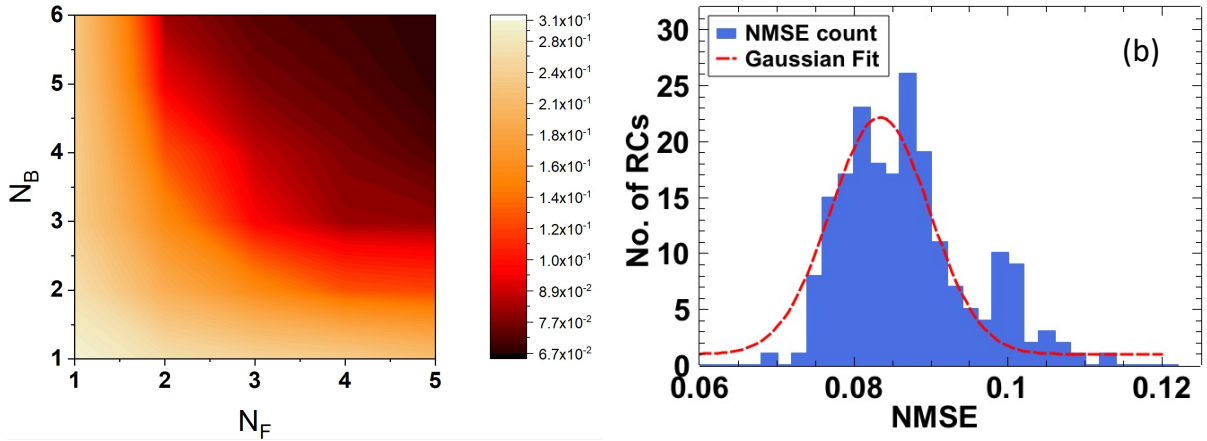

**Fig. 2. Results of NARMA-10 reproduction.** a) NMSE as a function of the number of MRRs per bank ($N_F$) and the number of filter banks ($N_B$) in ROSS-NN configuration. The actual number of spatial nodes is equal to $N_F \times N_B$. b) Distribution of NMSE values for random implementations of ROSS-NN internal connectivity in the form of random frequency offset of all filters' central frequency.

In this work, we explore a novel recurrent photonic integrated node consisting of a hardware-friendly filter-in-a-loop architecture that harnesses computational efficiency in a two-fold manner. First, the proposed architecture implements spectral slicing of the incoming signal through a complex node response, directly in the optical domain, leading to spectral decomposition of the signal, which is a prerequisite when broadband optical signals (from 100 GHz to several THz) are to be processed. Spectral slices consist of lower-bandwidth components of the original signal offering the possibility of diverse and specialized treatment of information in the frequency domain. Second, an optical filter provides a non-linear mapping of the incoming signal's phase variations to the node's intensity [20]. Hence, the coherent interaction of spectrally sliced components on the photodiode provides a complex nonlinear activation function at the output. This filter-based neuromorphic assumption of nodes is inspired by the filter-and-fire model [21] which treats each retinal ganglion cell as a linear filter followed by a nonlinear activation function. Spectral slicing, through simple non-recurrent filters, before RC processing was first proposed in [22] where the output of each filter was sent to a digitally implemented reservoir computing network. Nonetheless, this approach is not advantageous in terms of power consumption. Here, we implement a fully photonic structure based on filters-in-a-loop. Based on the functionality of the photonic node, we call the proposed architecture Recurrent Optical Spectrum Slicing Neural Network (ROSS-NN). Each node of the ROSS-NN exhibits truly passive operation, and there is no intrinsic bandwidth limitation burdening the proposed scheme. Processing speed is capped only by the bandwidth of the photodiodes and digital processing requirements in the form of linear regression at the system's output. A ROSS-NN can be incorporated either as an RNN or spatial RC architecture, providing direct coherent processing without costly electro-optic conversions, speed processing penalty, high-speed pseudo-random generators for mask realization [13] and most importantly, with marginal power consumption. Aiming to demonstrate the merits of ROSS-NN, we have numerically investigated its processing capabilities in two tasks. First, we confirm its nonlinear processing nature by exhibiting its efficacy in generic non-linear tasks, such as inferring the behavior of unseen data of a dynamical system using multi-Gsymbols/sec rates. Second, and more importantly, we also numerically demonstrate its performance in real-life problems such as the mitigation of transmission impairments caused by CD and Kerr effect in IM-DD systems at 112 Gbaud PAM-4 as well as in coherent systems employing QAM-16 signals. ROSS-NN exhibits uncontested superiority over well-established techniques such as Maximum Likelihood Sequence Estimation (MLSE) and Volterra Non-Linear Equalizers and other photonic neuromorphic approaches, extending the reach of high-speed IM-DD systems to long distances. The BER achieved by ROSS-NN is only limited by noise, as in typical linear channels.

**ROSS-NN node and overall architecture.** The basic unit of our system is a recurrent node consisting of a first-order bandpass or bandstop optical filter, two couplers and a feedback loop with delay $T_{d,..}$ The feedback loop is equipped with a phase shifter so as to adjust the feedback phase, whereas feedback losses ($L$) can be adjusted during fabrication or through the optional inclusion of a variable optical attenuator (VOA) (fig. 1a) [23]. The whole architecture can be monolithically integrated using mature silicon photonics technology, whereas the optical filters in the loop may be implemented by means of Mach-Zehnder Delay Interferometers (MZDI), Micro-Ring Resonators (MRR) or any equivalent bandpass/bandstop optical filter. In fig. 1b, we present a generic architecture consisting of multiple recurrent optical filters organized in separate filter banks, spectrally slicing different frequency bands of the input optical signal. This complex architecture can be easily implemented if add/drop MRRs are used, as rings can be interconnected using through ports and provide outputs using drop ports which are directed to the output layer. The output layer could be implemented in the optical domain, by combining filter's outputs through an optical combiner, followed by a single photodiode and ADC [24]; or in the digital domain with the use of a photodiode/ADC per filter output. Depending on the problem to be solved, the architecture may contain one or more filters incorporated in one or multiple loops. The number of filters or loops is mostly limited by optical losses and the corresponding signal-to-noise ratio (SNR) at the output layer. Each recurrent node focuses on a specific frequency band of the input optical signal. Thus, the number of nodes is on one hand related to the required granularity of spectrum slicing as dictated by the problem and should on the other hand be sufficient to properly cover the full optical bandwidth to be processed.

The transfer function of the recurrent node in fig. 1a is given by:

$$H_{node}(f) = \frac{\sqrt{1-a}\sqrt{1-b}H(f)}{1+\sqrt{a\,bL}\,H(f)\,e^{-i(2\pi fT_d+\varphi)}} \quad (1)$$

where $a$ and $b$ are the coupling ratios at the input and output, $L$ the VOA induced losses, $T_d$ is the total delay of the loop. $H(f)$ is the transfer function of the in-loop filter(s) and $\varphi$ is the phase imposed by the in-cavity phase shifter. The aforementioned nodes can be considered as building blocks of ROSS-NN which may serve as a photonic RNN or an RC. In particular, we have the possibility to follow the RC paradigm and mimic random inter-node connectivity by stochastically varying the complex amplitude of the signal injected from filter to filter in fig. 1b. To further enforce random connectivity, we can induce arbitrary frequency offsets between adjacent nodes that contribute to a stochastic mixing of the frequency components handled by successive filter nodes belonging to the same bank. In an RC-like treatment of the configuration depicted in fig. 1b, we follow the RC related training, thus restricting training only at the output layer of fig. 1b. On the other hand, one may handle all these variables (filter bandwidth, offset between successive filters, phase shifter, feedback attenuation etc) as hyper-parameters that can be optimized for a specific task. In this case, the network mostly resembles an RNN configuration where optical weighting between units can be applied in different forms (variation of signal amplitude and phase after each filter, variation of frequency offset between adjacent filters). Fig. 1c depicts the scheme that has been numerically tested and that provides remarkable results in the most important application of transmission impairments mitigation at extremely high baud rates (>100 Gbaud) and at low complexity. It will be shown in the results section that this simple scheme relying on two passive recurrent optical filters for PAM-4 and three filter for 16-QAM has the ability to vastly outperform state-of-the-art digital equalizers, while its complexity and therefore its energy footprint is orders of magnitude lower.

**Results**
**ROSS-NN for the NARMA task.** One of the key properties that a recurrent neuromorphic scheme should be able to address is the reproduction or prediction of pseudo-chaotic sequences with increased temporal complexity. Although these tasks (NARMA, Santa Fe, Mackey-Glass etc.) are of minor importance application-wise, their successful processing can assess the overall efficiency of a neuromorphic dynamical scheme. We chose the NARMA task, originally introduced in [25]. In this context, we utilized a pseudo-random input, drawn from a uniform distribution $u(n)$ and computed the tenth order NARMA-10 sequence $y(n)$.

$$y(n+1) = 0.3 \cdot y(n) + 0.05 \cdot y(n)[\textstyle\sum_{i=1}^{10} y(n-i)] + 1.5 \cdot u(n-10) \cdot u(n) + 0.1 \quad (2)$$

Each value from the pseudo-random input sequence $u(n)$ is used to modulate the amplitude of a laser source at a rate of 40 GSymbols/sec. The optical input is equally split and injected to a ROSS-NN configuration consisting of one to six banks ($N_B$), where each bank embeds one to five filters ($N_F$) that are numerically implemented as add/drop MRRs. Intra-cavity losses and the coupling coefficient between the circular and straight waveguides of the MRRs have been used as hyperparameters to tune the Q-factor and bandwidth of each MRR. The central frequency of each filter bank can be easily adjusted in a course way by placing a phase tuner inside the MRR. The drop ports from all filters are considered as the scheme's optical outputs and are fed to a typical detection scheme (photodiode and ADC) followed by a digital linear regression as the RC's trainable output layer. The regression has 10 taps, matching the NARMA's order. The purpose of the system is to train the output layer of the RC so that the system correctly emulates the sequence $y(n+1)$ after training. To achieve this, 50% of the NARMA outputs of 4000 symbols was used as a training sequence, regulating the weights of the linear regression. Following the training procedure, the RC was fed with 2000 $u(n)$ subsequent symbols and the generated output $\hat{y}(n+1)$ was recorded. Accuracy was evaluated by computing the normalized mean square error (NMSE) between $y(n+1)$ and $\hat{y}(n+1)$. Taking into account that ROSS-NN is based on the spectral slicing property, for each combination of banks/filters, all the critical parameters such as the MRR's $Q$-factor, the detuning of each bank relevant to the signal's bandwidth and the central frequency of each MRR resonance compared to the banks center, were scanned so as to achieve optimum performance. In fig. 2a, it can be seen that in order to get normalized mean square error (NMSE)<0.1, a minimum of number banks equal to $N_B$=3 each having $N_F$=4 MRRs is needed, resulting to only 12 physical nodes. A critical observation is that for each neural configuration ($N_F$, $N_B$) each filter's bandwidth and each bank's spectral band is optimized so that the full spectrum of the incoming signal is covered. Up to now we did not really treat the ROSS-NN as a strict RC network. Although we restricted training in the linear regression part, we also tried to optimize all the other hyperparameters related to the number of filters per bank, the number of banks and the exact shape in terms of bandwidth and central frequency of each individual MRR. This treatment contradicts with one of RC's most fundamental aspects, the randomness of connections, that contributes to its hardware friendliness. In order to evaluate the impact of randomness on performance, we have assumed realistic structural deviations, as if this scheme was realized in a typical silicon photonic platform [26, 27] (methods). In particular, we solved the same NARMA task using 200 ROSS-NN instances (RCs in fig. 2b) each having $N_B$=5, $N_F$=5. These ROSS-NN instances exhibit structural deviations compared to an "ideal" prototype, such as effective refractive index of each MRR due to waveguide roughness, resulting to frequency detuning deviations, inter-MRR transmission coefficient and delay (phase). The "ideal" in our case is a ROSS-NN whose key properties (detuning, bandwidth) are optimized for the specific task as hyperparameters and no fabrications related imperfections are considered. In fig. 2b, the histogram for this scenario alongside a gaussian fit is presented showing that the $\overline{NMSE} = 0.086 \pm 0.0005$, not deviating significantly from the "ideal" NMSE=0.079. The NARMA results obtained from the ROSS-NN scheme can be directly compared to other RC realizations that offer an NMSE=0.168 using 51 time-multiplexed nodes [28]. Hence, while preserving marginal power consumption and integration capabilities, ROSS-NN offers ≈50% reduction in node count compared to the state of the art. Furthermore, the proposed scheme can address this task without any speed penalty that is present in time-delayed RCs, whereas even higher bandwidths can be envisioned without any additional considerations apart from SNR, the elevated analog bandwidth of the photodiodes and the ADC, which are typical limitations for all photonic neuromorphic or general signal processing schemes.

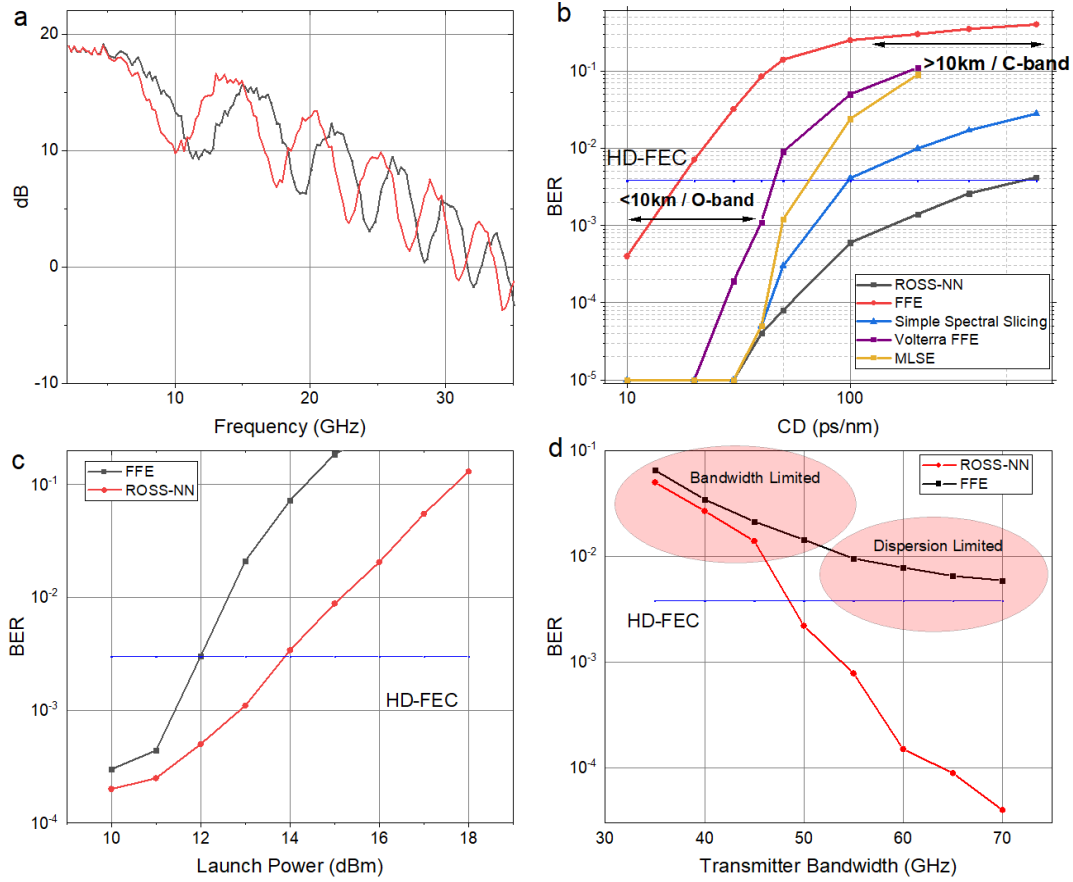

**Figure 3: Performance of ROSS-NN in the mitigation of transmission impairments in intensity modulated optical communication links**. a) The spectral response of the recurrent nodes after photodetection. The power fading effect, due to 20 km C-band transmission, is observed with multiple spectral dips in both outputs, however frequency diversity is also observed. With the proper adjustment of bandwidth, frequency detuning and delay values characterizing the ROSS-NN system, we can almost completely eliminate power fading and provide additional memory for tackling intersymbol interference caused by CD. b) The CD tolerance of the proposed RC-system in comparison with a system exploiting two filters without any feedback and FFE, MLSE, VNLE as post-processing. While for small accumulated CD values (for example <20km reaches in the O-band), two simple filters or heavy MLSE, VNLE can achieve acceptable results, for higher CD values, only the system with the recurrent nodes achieve results below HD-FEC limit. c) The efficiency of the proposed system in the mitigation of Kerr-related non-linearities when CD is optically compensated. The system provides 2 dB gain in comparison with a linear algorithm. d) The performance of the ROSS-NN-system and the FFE as a function of transmitter's bandwidth for fixed 35 GHz bandwidth per photodiode. The results refer to a 40 km O-band link with a GVD parameter D=0.5 ps/nm km.

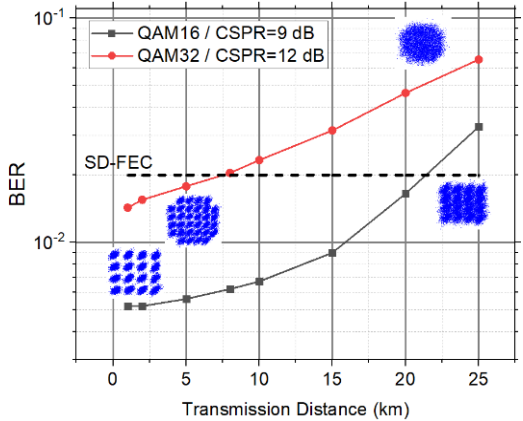

**Fig. 4**. **BER results as a function of transmission reach for 16-QAM and 32-QAM in the self-coherent configuration.** With the QAM-16 format, 20 km of transmission is achieved, while with QAM-32 the reach is at least 5 km.

**Few-node ROSS-NN as a photonic hardware accelerator in 100 Gbaud and beyond optical communication systems.** Until today, data center interconnects are mainly based on cost-effective direct detection systems covering distances from 500 m to 80 km. The main limitation in such distances is the interaction between CD and the square-law of the photodiode which results in power fading. Generically, a real-valued unipolar signal at the transmitter with a direct current (DC) bias can be expressed as:

$$r(t) = s(t) + c(t) \quad (3)$$

in which $s(t)$ is the original signal and $c(t)$ is the optical carrier related to DC bias. The received signal with square-law detection is represented as:

$$y(t) = c^2(t) + |s(t) \otimes h(t)|^2 + 2 c(t) \cdot s(t) \otimes F^{-1}[Re\{H_{fiber}(f)\}] \quad (4)$$

where $\otimes$ is the convolution operator, and

$$Re\{H_{fiber}(f)\} = cos(2\pi^2 \beta_2^2 L f^2) \quad (5)$$

in which $\beta_2^2$ is the second order dispersion coefficient, $L$ denotes fiber length and $f$ represents signal frequency. Based on Eq. (4), it can be found that the received signal suffers from power fading caused by CD and its nonlinear transformation at the photodiode. This dispersion-induced power fading will result in deep spectral zeros when $2\pi^2\beta_2^2Lf^2 - \pi/2$ is a multiple of $\pi$.

Many works in the literature have been devoted to the mitigation of this distortion and a number of techniques such as optical dispersion compensation, Single Sideband (SSB) modulation, digital equalization in the form of DFE [29,30] or Maximum Likelihood Sequence Detectors (MLSD) [31, 32] have been reported. The quadratic dependence of CD on baudrate is the reason why next-generation 112 Gbaud IM/DD links are forced to rely on heavy digital signal processing (DSP) algorithms that cancel out accumulated dispersion up to 10 km, while for longer links, coherent detection is the only viable, but expensive solution. ROSS-NN is capable of equalizing both IM/DD and coherent schemes. We first demonstrate that ROSS-NN consisting of two nodes is capable of mitigating transmission impairments in an IM/DD at unprecedented baudrates (> 112 Gbaud) achieving even 60 km reach with very high CD tolerance. In general, an intuitive way to understand RC operation, is as a nonlinear dynamical system that acts as a pre-filter on the input data, transforming it into a higher dimensional space [33]. This is achieved by using a transformation resulting in multiple outputs which have undergone different routes in the spatial, temporal and - most importantly in our case – the spectral domain. The recurrent connectivity offers rich and frequency-dependent memory (see supplementary information) which is important when transmission impairments are caused by nonlinear channels with memory, such as single-mode fibers. The proposed ROSS-NN consisting of two recurrent nodes (fig. 1c) provides evident frequency diversity of power fading, characterizing the distorted signals at the two outputs due to CD (see fig.3a). This is achieved by treating differently the lower and higher frequency components through spectral slicing of each sideband and by leveraging optical feedback as an extra mechanism to enhance specific frequency components and fading memory. Both outputs are followed by photodiodes with bandwidth lower than the baud rate and ADCs that require only one sample per symbol, thus showing that real-life implementation of the scheme is practical at high baud rates (>100 Gbaud) and more appealing than coherent detection which requires at least 1.25 samples/symbol in order to decode the signal [34]. An FFE follows the ADC in order to act as a linear regression stage and to assist in the elimination of Inter-symbol Interference (ISI) and bandwidth limitations caused by CD and transceiver opto-electronic components. As for the accumulated CD tolerance, fig.3b compares the proposed system with a system consisting of two simple optical filters offering frequency diversity [22] and an FFE, showing that, especially for high CD values as the ones in C-band, and for reaches beyond 10 km, ROSS-NN is the only viable solution with one order of magnitude better BER performance. Furthermore, the ROSS-NN outperforms state-of-the-art digital algorithms like MLSE with 5 taps and a 3rd order Volterra Non-Linear Equalizer (VNLE) with 91,31,11 taps for each order (fig. 3b). ROSS-NN compared to VNLE can perform superior equalization, requiring only 40-100 multiplications, whilst Volterra requires over 2400 for the results presented in fig. 3b.

In order to benchmark the RC-system in a harsh nonlinear transmission environment, we compensate CD using a dispersion compensating fiber (DCF) with a high non-linear parameter and launch powers in the numerical model that excite Kerr non-linearities. By comparing ROSS-NN with a simple FFE as a post processing method, 2 dB higher tolerance in nonlinear effects is achieved (fig. 3c). ROSS-NN is also capable of mitigating bandwidth limitations induced by optoelectronic components of the transceiver. The bandwidth provided by vendors of Mach-Zehnder modulator (MZM) drivers and Digital-to-analog converter (DAC) is in the order of 60 GHz, which poses strict limitations to the extension of baud rate beyond 112 Gbaud. Although FFEs and Pre-Emphasis filters are recognized tools for the mitigation of this distortion, it still constitutes a major problem. In fig. 3d we provide results for the tolerance of the proposed system to the limited bandwidth of the transmitter. The photodiode bandwidth after each node is assumed constant at 35 GHz. It is shown that even with 50 GHz analog bandwidth in the transmitter, sub-HD-FEC results could be achieved even for a 40 km long O-band link. If the 25% overhead SD-FEC is considered, the analog bandwidth could be reduced to almost 45 GHz. It must be stated that all critical hyperparameters of the ROSS-NN have been optimized in this study (see supplementary information).

We further benchmarked a three-node system in a coherent transmission link in order to prove the versatility of the ROSS-NN and its suitability to deal with coherent modulation formats. We take advantage of a residual carrier that permits the reception of the coherent signal with simple PDs following the paradigm of cost efficient self-coherent systems [35]. State-of-the-art 120 Gbd QAM-16 and QAM-32 scenarios are numerically simulated solely focusing on chromatic dispersion mitigation with the use of ROSS-NN. In this baudrate, 400 Gbps and 500 Gbps net data rates can be achieved in a single wavelength and polarization. We keep the Carrier-to-Signal Power Ratio (CSPR) within the limits of a typical Kramers-Kroning (KK) receiver [35], namely between 9 and 12 dB. In such systems, the CD effect is linear if coherent detection is utilized. In this work we choose the much simpler direct detection leveraging the residual carrier, however dispersion management without using the efficient but computationally heavy KK algorithm, constitutes a rather challenging signal processing problem. By employing ROSS-NN, we perform phase-to-amplitude conversion that maps all the different QAM symbols to the amplitude domain. Spectral slicing by three nodes also relaxes the need for large analog bandwidth in the order of 45-50 GHz, thus constituting an extremely attractive solution in the bandwidth-hungry area of coherent communication technology. The readout is split in two linear layers, one per quadrature. In fig. 4, indicative results of the BER performance of the two modulation formats is presented as a function of transmission distance in an O-band link. Sub SD-FEC performance is achieved after 20 km using 16-QAM with a CSPR of 9 dB while almost 8 km reach is achieved with 32-QAM and CSPR of 12 dB. This is the first time that such a simple direct detection scheme suited for M-QAM at high baud rates is presented practically based on an extremely simple DSP at the back-end of the receiver. It must be also stated out that the BER tolerance to transmitter phase noise is very high (see supplementary information). The results depicted in fig. 4 consider transmitter linewidth in the order of 100 kHz, whilst coherent receivers require narrow linewidth lasers (< 30 kHz).

**Discussion**

In this paper we have proposed and numerically evaluated a new neuromorphic photonic concept based on recurrent optical spectrum slicing; implemented by optical filters embedded in a delay loop. Such a concept constitutes a practical and realization-ready solution in silicon photonics chips [36] or even leveraging programmable photonics platforms [37]. The main advantage of the proposed scheme is its compatibility with direct processing in the optical frequency domain, thus rendering the specific neuromorphic approach suitable for spectral decomposition and processing of ultra-broadband signals (~ THz). Especially, when ultra-fast processing is necessary, a solution that can easily scan and process broad optical spectra directly in the optical domain and with minimum power consumption or need for data storage is a really useful tool. ROSS-NN can cover this need and play the role of a high-speed optical frequency processor in applications such as high throughput real-time flow-cytometry [38], high-resolution 3D imaging [39] and in general in tasks where simultaneous spectro-temporal knowledge is required at very fast rates.

Since optical communications industry has the strongest foothold in photonic applications for real-life problems, we anticipate that the impact of ROSS-NN in the advent of edge-cloud interconnects will be significant. Edge-cloud era seeks for straightforward, low-cost but novel ideas for facing the extreme requirements in low latency, high bandwidth, stability and power efficiency. Already, moving digital processing as near as possible in the optical transceiver, through co-packaged optics is a colossal migration step, which will disrupt the field in the next decade [40]. But, implementing computing and processing directly in the optical domain, in the core of the optical engine, is an ambitious endeavour. By relaxing the optical bandwidth requirements (less than 50 GHz opto-electronic components for 100 Gbaud and beyond signals as shown in fig. 3, 4) and keeping DSP to the bare minimum, through optical pre-processing, ROSS-NN could counteract the severe power consumption issues that 800G technology poses. In comparison with coherent technology, which conquers even the shortest reach scenarios of Inter Data center Communications [41], ROSS-NN could offer multi-Watt reduction in the overall transceiver power budget, either in the IM/DD or in its self-coherent approach. Combating CD even in Extended Reach (ER) 40 km channels with uncontested energy consumption, ROSS-NN receivers constitute an appealing tool for the 6G, Internet-of-Everything and Industry 4.0 revolutions of the next years.

An interesting field towards further exploration of ROSS-NN systems is their training. As already pointed out, ROSS-NN can be used as building block for both RC and RNN implementations. In the former case, the readout layer is one of the most critical parts of the architecture. Digital or even optical readout should be studied in-depth in forth-coming studies. First evaluations show that optical readout further enhances the performance of the network in specific tasks from the telecom arena, due to the fact that the nonlinear activation is boosted when all favored frequency components from diverse nodes are combined on the same square law detector [24]. When the network is operated as an RNN, then activities on training become even more demanding as all hyperparameters along with the readout layer must be optimized concurrently utilizing back-propagation or equivalent techniques. Another critical property of ROSS-NN attributed to its recurrent nature is that ROSS-NN has the property to enhance the number of output traces through time multiplexing which is equivalent to temporal unfolding of each spatial node's dynamical behavior [42]. More complex networks can be realised if the characteristic delay of the loop of each bank is varied. Definitely, ROSS-NNs open new paths in investigating and training neural networks in the frequency domain and could be also considered as a neuromorphic approach even in the electronic domain where the implementation of analog filters with diverse transfer functions is mature and CMOS technology permits the hardware implementation of complex networks consisting of thousands of filter nodes with unprecedented granularity.

**Methods**

**Recurrent node simulation.** In this work, we propose a novel recurrent filter node for RNN/RC architectures. Such recurrent filters are easily integrated into photonic circuits along with many other optical components like semiconductor optical amplifiers (SOAs), VOAs, phase shifters, couplers etc. [43,44]. In our simulations, each filter is modelled through its transfer function $H(f)$, while the phase shifters, VOAs are inserted as phase and feedback terms in (1).

The transfer functions of the MZDI and MRR filters $H(f)$ are given by

$$H_{MZDI}(f) = \frac{1}{2}\left[1 + e^{-i2\pi(f-f_0)\Delta T}\right] \quad (7)$$

where $f_0$ is the central frequency of the filter and $\Delta T$ is the delay difference between the two arms. $f_0$ can be tuned with the use of phase shifter in one of the two MZDI arms. Regarding MRRs, the through port and the drop port transfer functions are given by (8)-(9),

$$H_{MRR,Through}(f) = \frac{-T_2 e^\Phi + T_1 - (K_1^2\, T_2\, e^\Phi)}{(1 - T_1 T_2 e^\Phi)} \quad (8)$$

$$H_{MRR,Drop}(f) = \frac{-K_1 K_2\, e^\Phi}{(1 - T_1 T_2 e^\Phi)} \quad (9)$$

$$\Phi = \frac{-a\, L/2 - [i2\pi(f-f_0)\, L\, n_{eff}]}{c} \quad (10)$$

where $T_1$, $T_2$ the transmittance, $K_1$, $K_2$ the coupling coefficientσ, $a$ the waveguide losses, $n_{eff}$ the effective refractive index, $L$ the circumference of the ring and $c$ the speed of light.

**Transmission system simulation – ROSS-NN evaluation in optical communication tasks.** The transmission system consists of a semiconductor laser modelled with the well-known Lang-Kobayashi rate equations [45] for the complex slowly varying amplitude of the electrical field $E(t)$ and the carrier number inside the cavity $N(t)$.

$$\frac{dE}{dt} = \frac{1+i\alpha}{2}\left[G_s - \frac{1}{t_{ph}}\right]E_f + \sqrt{2\beta N}\xi \tag{11}$$

$$\frac{dN}{dt} = \frac{I}{q} - \frac{N}{t_n} - G|E|^2 \tag{12}$$

$$G = \frac{g[N-N_0]}{1+s|E|^2} \tag{13}$$

TABLE I
Numerical Model Parameters

| Symbol | Parameter | Value |
| --- | --- | --- |
| $g$ | Differential gain parameter | $1.2 \times 10^{-8}$ ps$^{-1}$ |
| $s$ | Gain saturation coefficient | $5 \times 10^{-7}$ |
| $\beta$ | Spontaneous emission rate | $1.5 \times 10^{-10}$ ps$^{-1}$ |
| $t_n$ | Carrier lifetime | 2 ns |
| $N_0$ | Transparency Carrier Number | $1.5 \times 10^8$ |
| $A$ | Linewidth enhancement factor | 3 |
| $\omega_0$ | Central oscillation frequency | $1.206 \times 10^{15}$ rad/sec |
| $I$ | Bias current | 35 mA |

Here $\alpha$ is the linewidth enhancement factor, $g$ is the gain parameter, $s$ is the gain saturation coefficient, $t_{ph}$ is the photon lifetime, $t_n$ is the carrier lifetime and $N_0$ is the carrier number at transparency.

The symbols for the laser modulation rely on the Mersenne Twister pseudo-random generator with a unique seed and a repetition period of $2^{19937}-1$. The reason for this measure is to hinder ROSS-NN from anticipating the next symbol in the sequence and thus, overestimate the equalization results. An external MZM is assumed, acting as a 2$^{nd}$ order Butterworth filter, emulating bandwidth limitation at the transmitter. We simulate, with the integration of Nonlinear Schrödinger equation (NLSE) using the Split-step Fourier method, the transmission of 112 Gbaud PAM-4, QAM-16 signals in a range of 10 km to 60 km transmission distances. Signal propagation in our model is governed by Manakov equations [46]. The GVD parameter takes values between $D=0.5$ ps/nm km and $D= 4$ps/nm km for O-band transmission, while $D=17$ ps/nm km is assumed for simulations in the C-band. The fibre losses are set to $a=0.34$ dB/km in O-band and $a=0.21$ dB/km in C-band. The non-linear parameter is $\gamma=1.3$ W$^{-1}$ km$^{-1}$, while when DCF is used, its $\gamma=6$ W$^{-1}$ km$^{-1}$. In the receiver side, a pre-amplifier with 5 dB noise figure is simulated in order to compensate for the transmission losses, the chip's insertion loss and the initial splitter. In a real-life scenario, a semiconductor amplifier in single wavelength transmissions or a Doped Fiber Amplifier in wavelength division multiplexing scenarios could play the role of pre-amplifier. The intensity of recurrent nodes output is captured with PDs modelled as a square-law element of responsivity $R=0.8$ A/W and bandwidth 35 GHz. Shot and thermal noise are taken into account. An 8-bit, 112 Gs/s ADC follows each PD, with analog bandwidth of 35 GHz.

**Training of the readout layer.** The ADC generates one sample per incoming symbol. These digital samples are inserted to a linear classifier that resembles the typical symbol-spaced FFE block in IM/DD DSP. The length of the FFE is adjusted so as to match the channel's memory which is proportional to the group delay time $T=D*\Delta\lambda*L_D$, where $D$ is the second order dispersion parameter, $\Delta\lambda$ the optical bandwidth occupied by the signal and $L_D$ the transmission distance. In these simulations the memory ranges from 11 to 21 symbols for O-band transmission, while for C-band links this number reaches up to 71 symbols. Half of the symbols are considered as pre-cursor and half as post-cursor taps. The weights, $b$, of the linear equalizer are calculated by finding the pseudo-inverse matrix through Tikhonov regularization. With 20000 symbols for training and 100000 symbols for testing, we achieve enough precision for BER above 10$^{-4}$. When QAM is considered (fig. 4), we apply two separate linear readouts, one for the real part and one for the imaginary part (see supplementary information).

**NARMA10 task.** The pseudo random signal that is used as input for the NARMA sequence consists of 4000 samples drawn from a uniform distribution (python's generator). The values range from 0 to 0.5. The NARMA output is computed assuming to have an order (memory) of 10. Simulation wise, the pseudo random values are oversampled using 8 samples per symbol and the time scale was regulated so as to result to a rate of 40Gsymbol/sec. These analog values were used to amplitude modulate a continue-wave (CW) laser with power of 0 dBm, assuming an extinction ratio of 20 dB. The optical signal was assumed to be amplified, using an amplifier gain of 10 dB and a noise figure of 5 dB. The signal was subsequently split according to the number of ROSS-NN nodes used. In this case inside each ROSS-NN module, add/drop MRR filters are assumed. The outputs from the MRR's drop ports were recorded by PDs and typical shot and thermal noise was incorporated. The computed photocurrents were normalized and they were fed to a linear regression algorithm with 10 taps so as to match the NARMA's memory. 2000 samples of the NARMA were used as teacher so as to train the weights of the linear regression. Following this step, 2000 samples from NARMA's pseudorandom input were fed to the ROSS-NN for inference; aiming to reproduce the actual NARMA output. The two traces (predicted and reproduced) were compared using the normalized mean square error (NMSE).

Regarding the neuromorphic architecture used for addressing NARMA, we varied the number of banks (ROSS-NN node) and MRRs per bank. For all instances the MRRs were assumed to have a radius of 55 μm and propagation losses of 0.4 dB/cm. The waveguides connecting the MRRs per bank were assumed to exhibit transmission coefficient of 0.95, whereas inter-MRR delay was fixed at 0.1 of symbol duration. The detuning of each MRR was assumed to be such that the combination of banks and filters span over the whole bandwidth of the signal. Therefore, the center frequency of each MRR and the spacing among different filters was scanned for each combination of banks and filters per bank. In the same context, the coupling coefficient for each MRR, partially regulated filters bandwidth and was scanned so as to locate the lower NMSE during inference. The delay in each bank was set to 1 symbol time and feedback strength was set to 0.5.

RC random synapses realization: so as to evaluate the impact on NMSE of parameter deviations, we fixed the number of banks and filters ($N_F$=5, $N_B$=5) and optimized all the other parameters so as to locate the lowest NMSE for this setup. This optimized neural network was considered as "ideal". Then we generated 200 RC instances where all the parameters randomly varied using a uniform distribution, with a range of +/- 10% with respect to the ideal. The parameters subject to this perturbation were: the center frequency of each MRR and the inter-MRR transmission efficiency. In addition, following [27,28] we assumed that for the inter-MRR connections, the effective refractive index of each waveguide varies following a normal distribution with standard deviation of $\Delta_{neff}$ =0.15 due to roughness.

**Acknowledgements**
The work has received funding from the EU H2020 NEoteRIC project under grant agreement 871330 and by the Hellenic Foundation for Research and Innovation (H.F.R.I.) under the "2nd Call for H.F.R.I. Research Projects to support Faculty Members & Researchers" (Project Number: 2901)

**Author contributions**
All authors contributed extensively to the work presented in this paper and to the writing of the manuscript.
**Competing interests**
The authors declare no competing interests.
**Additional information**
Supplementary information

# High Speed Photonic Neuromorphic Computing Using Recurrent Optical Spectrum Slicing Neural Networks: Supplementary Material

K. Sozos[1], A. Bogris[1], P. Bienstman[2], G. Sarantoglou[3], C. Mesaritakis[3]

[1]Dept. of Informatics and Computer Engineering, University of West Attica, Aghiou Spiridonos, 12243, Egaleo, Greece

[2]Dept. of Information Technology, Ghent University-imec, Technologiepark Zwijnaarde, 9052 Gent, Belgium

[3]Dept. Information and Communication Systems Engineering, Engineering School, University of the Aegean, Palama 2, 83200, Samos, Greece

This document provides supplementary information to "High Speed Photonic Neuromorphic Computing Using Recurrent Optical Spectrum Slicing Neural Networks", including details on the transfer function characteristics of the ROSS-NN, the optimization of ROSS-NN in high baud rate telecom applications, its memory capacity as well as a more extensive conversation about the readout layer.

**A. ROSS-NN node transfer function characteristics**

The neuromorphic node is a described as a filter-in-a-loop followed by a nonlinear element (photodiode in this paper). The recurrent filter properties define the weights of the filter in the spectral domain and its nonlinear impact as the output of the filter that will be driven to a photodetector will finally provide intensity being equal to $P_{out}=|h_{node}(t)*x(t)|^2$ where $h(t)$ is the impulse response of the filter-in-a-loop related to the transfer function of (1) (the same as the one appearing in the paper, also presented here for the sake of completeness).

$$H_{node}(f) = \frac{\sqrt{1-a}\sqrt{1-b}H(f)}{1+\sqrt{a\,b}L\,H(f)\,e^{-i(2\pi f T_d + \varphi)}} \qquad (1)$$

$$h_{node}(t) = \mathcal{F}^{-1}[H_{node}(f)] \qquad (2)$$

Interesting spectral characteristics emerge from the transfer function of the filter, such as selectivity and unequal amplitude response for frequencies being symmetrically placed with respect to central frequency. The versatile frequency response of the filter-in-a-loop provides also interesting properties as far as memory capacity is concerned as it will be shown in a subsequent paragraph. The transfer function can be programmed by adjusting critical parameters such as feedback delay time ($T_d$), feedback strength (depending on $a$, $b$, $L$ parameters), the phase shift ($\varphi$) of the feedback loop and the frequency offset with respect to central frequency of the input signal. In fig.1, the transfer function of a node is plotted in different scenarios where each hyperparameter is varied. A variety of amplitude (and phase) responses can be obtained. For the sake of simplicity here we consider the transfer function of a first order Butterworth filter regarding $H(f)$. The bandwidth of the filter is set to 35 GHz. Apart from $T_d$ which is a relatively hard parameter with minimal tuning flexibility, the other parameters can be easily tuned for an implemented chip.

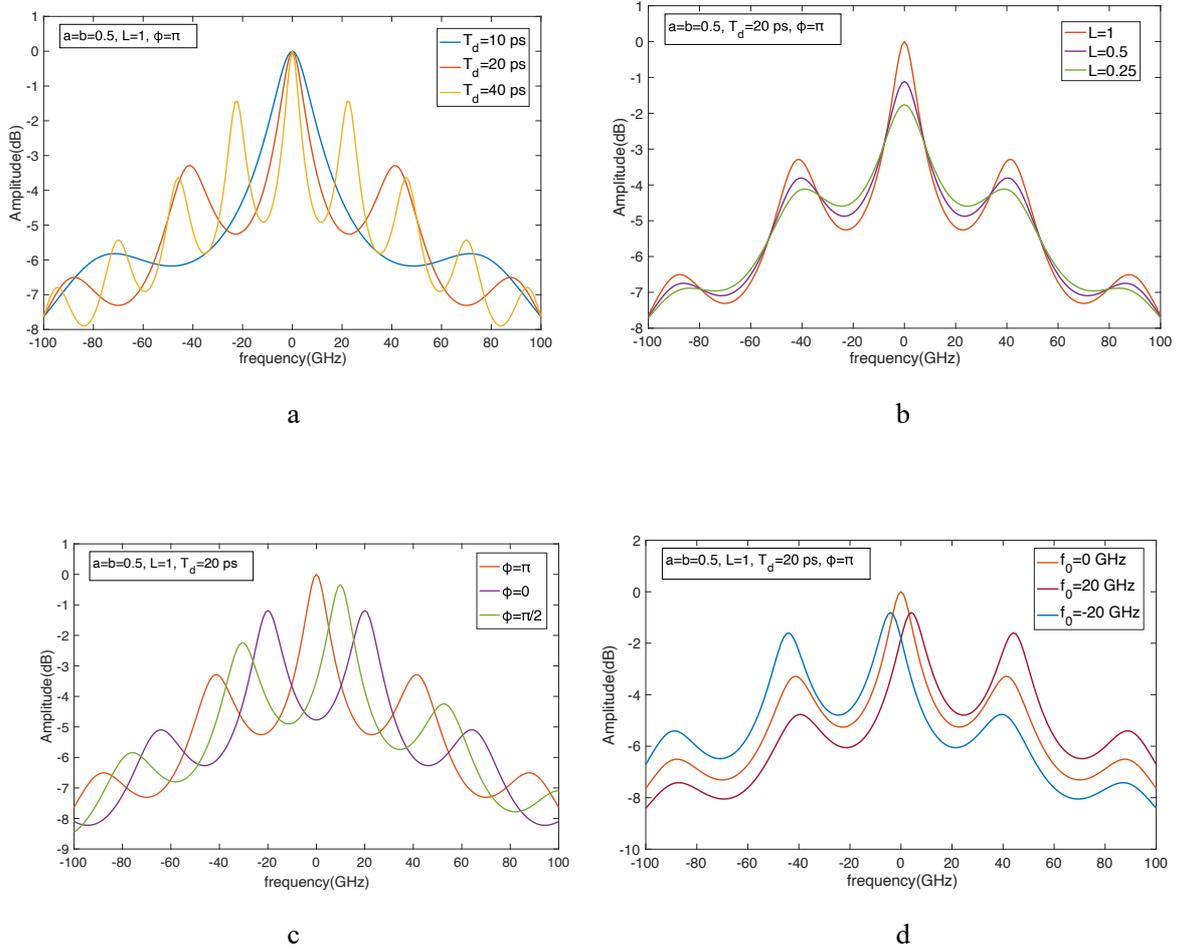

**Fig. 1. Node transfer function as a function of critical parameters.** a) Effect of time delay $T_d$, b) effect of feedback strength $L$, c) effect of phase shift $\varphi$, d) effect of frequency offset $f_0$ with respect to signal's central frequency.

## B. Readout Layer and hyperparameters role in the mitigation of transmission impairments

As mentioned in the manuscript, the linear readout layer after the filter nodes, could be implemented either with optical weighting and a single PD and ADC (optical readout), or with a PD and ADC pair after each node followed by a digital linear regression unit (digital readout). The two main approaches are depicted below.

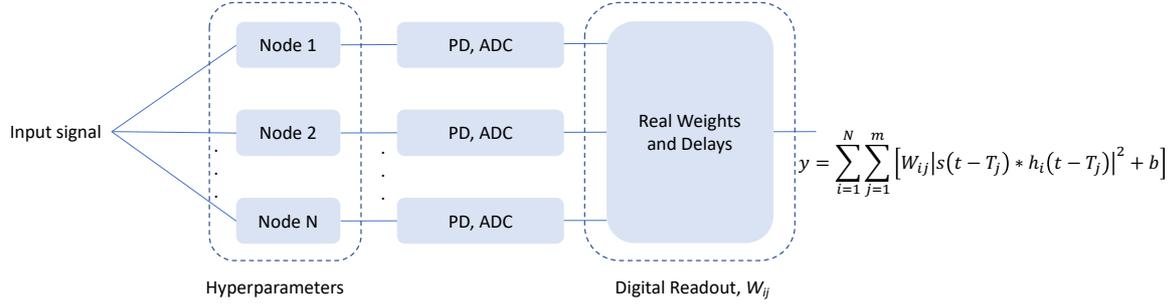

Fig. 2. Digital readout for real signals (PAM-4, NARMA)

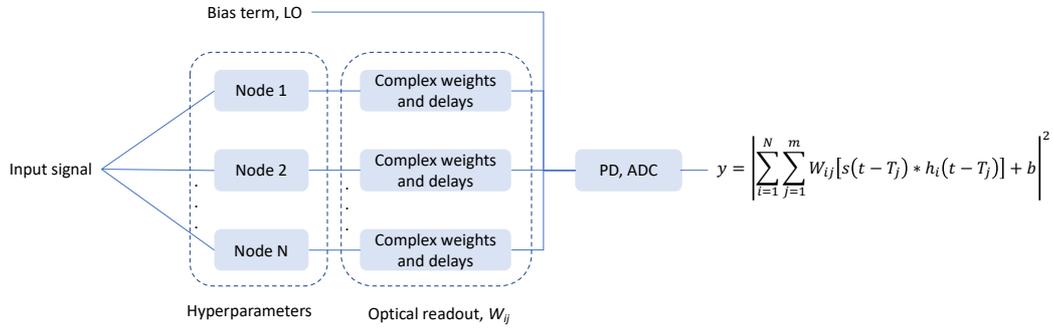

Fig. 3. optical readout

In the manuscript, we present results that rely on electrical readout depicted in fig. 2 which is more straightforward to implement in real-life. That is, the output of each recurrent filter is photodetected, sampled and then sent to a linear regression stage where we have also included memory at the resolution of a symbol so as to correlate past and future symbols with respect to the current one as a typical Feed Forward Equalizer (FFE) does. Therefore, if we have to include $k$ delays taps in the FFE, where $k=2m+1$, then the symbols $s_{i-m}$ to $s_{i+m}$ will be considered as inputs in the linear regression part per filter output. This approach is followed in dispersion mitigation problems presented for PAM-4, 16-QAM in the paper. The number of delay taps $k$ depends on the memory of fiber channel which is proportional to the group delay time $T=D*\Delta\lambda*L_D$, where $D$ is the second order dispersion parameter, $\Delta\lambda$ the optical bandwidth occupied by the signal and $L_D$ the transmission distance. In NARMA-10 benchmark we use only 10 taps for past symbols in the readout.

In figure 3, we depict optical regression based on [1]. This approach requires less opto-electronic and electronic components (PD, ADC) and as shown in the equations included in fig. 3, it enhances the coherent interaction of the spectral components coming from all nodes resulting in more complex nonlinear activation after photodetection. This is also reflected on the BER results, not shown in a graph, which present a significant improvement, exceeding an order of magnitude, compared to the ones presented in fig. 3 of the manuscript. However, optical readout requires coherent interaction with an optical bias which is equivalent to implementing coherent detection that is sensitive to phase noise and frequency fluctuations. In that case, extra DSP is required in order to attain carrier synchronization. Moreover, the training procedure of the optical weighting unit may be challenging, as extra delay lines equipped with phase shifters and variable optical attenuators are required. All the aforementioned difficulties are avoided taking advantage of digital readout using a few (two or three in high baud rate telecom systems) lower bandwidth receivers which reduce the necessity for high bandwidth operation for both PDs and ADCs.

When coherent signals are sent to the digital readout, then two readout units are trained, one per quadrature as depicted in fig. 4.

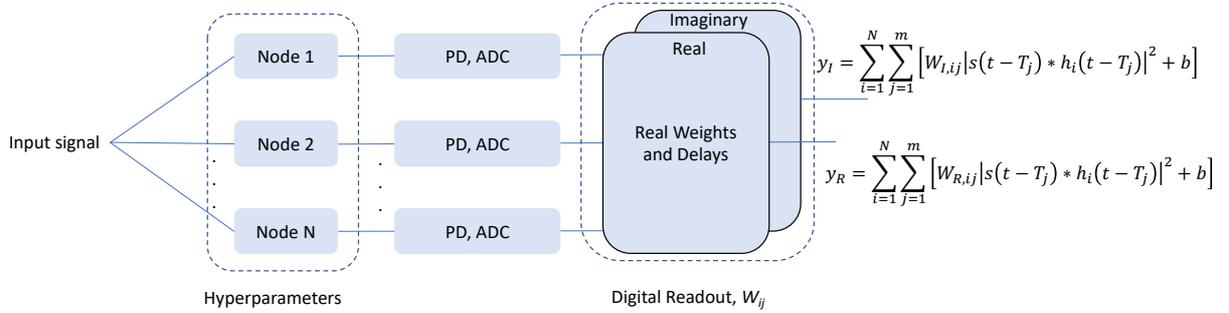

**Fig. 4. Digital readout for complex signals (QAM-4)**

In PAM-4 and QAM-16 transmission systems presented in the manuscript we perform a complete scanning of critical hyperparameters of each node, as the ones presented in fig. 1 of supplementary information, in order to identify regimes of better equalization performance. For instance, the desired diversity in Chromatic Dispersion (CD) mitigation was achieved through the versatile transfer function of the recurrent node which depends on the hyperparameters of filter bandwidth, frequency offset with respect to the central frequency of the incoming signal. Regarding the rest parameters, $L=1$, $a=b=0.5$ and $\varphi=\pi$, $T_d=9\ ps$. In fig.5, a cross investigation of the two important parameters, filter bandwidth and frequency offset, is carried out, illustrating the conditions under which the best differentiation and optimum performance is achieved when CD is the dominant impairment in 112 Gbaud PAM-4 system. A comparison is carried out between receivers using filters with and without feedback, showing the superiority of recurrent processing (as shown in left and right subfigures of fig. 5). More specifically, in the presence of feedback loop, the system transforms to a high-dimensional dynamical system and this is also evident in the BER performance (an order of magnitude improvement). These results were achieved considering simple Mach-Zehnder Delay Interferometers (MZDI) filters in the loop. In a real system, the bandwidth of the recurrent filter and the delay are hard parameters, not easy to tune. However, for a different bandwidth or delay parameter, there is a flexibility in acquiring very good performance if frequency offset is properly tuned as shown in fig. 5.

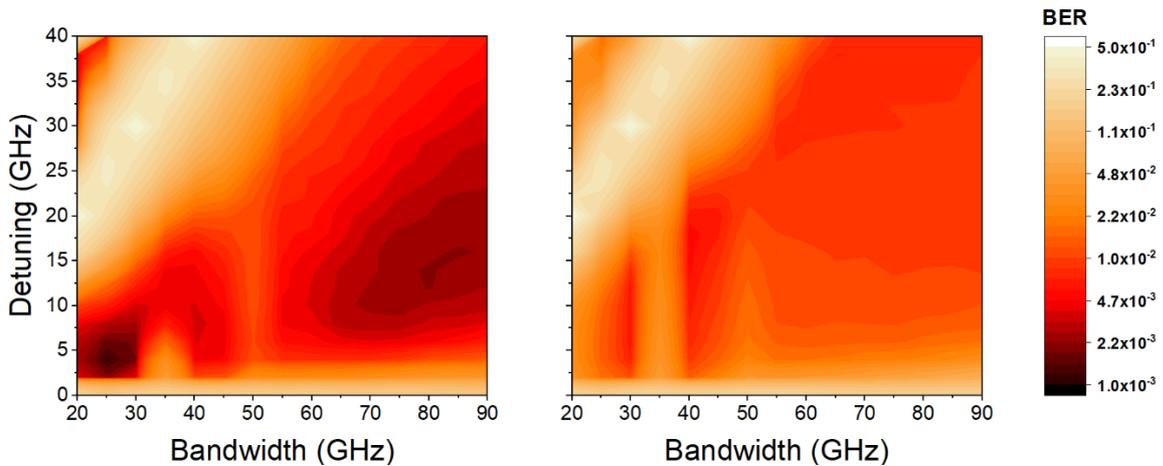

**Fig. 5. BER performance as a function of filter bandwidth and frequency detuning in respect to the carrier.** The benchmark is numerically carried out considering a 40 km O-band link with a GVD parameter $D=3$ ps/nm km. (Left) The proposed RC system with two recurrent nodes. The two filters are partially overlapping with acceptable results in a wide bandwidth area. For small filter bandwidth, a frequency detuning of 5 GHz is enough for good differentiation, while the greater the bandwidth the greater the flexibility with respect to inter-filter detuning. (Right) The system of two simple filters providing frequency diversity shows inferior results with a similar trend. The periodicity that is observed in the BER results is due to the periodical spectrum of the MZDI.

Regarding QAM-16 detection, our approach is an alternative to KK approach [2] that can extract both quadratures with the use of direct detection. Our scheme takes advantage of the strong residual carrier and recurrent filtering so as to differentiate the 16 or 32 symbols at the output. If one uses simple filters, it is not possible to extract the symbols efficiently, therefore the recurrent operation is indispensable in this example as well. The coherent transmission is performed in O-band with D=3 ps/nm*km and the optimization of the system is carried out based on filter frequency and inter-filter detuning (contour plot of fig. 6b). The rest of the parameters, such as the feedback Delay Time ($T_d$), Feedback Strength ($L$) and the Carrier-to-Signal-Power-Ratio are also optimized in extensive simulation trials. The role of the transmitter linewidth is also investigated and the results are presented in the fig. 6a. The relation between BER results and linewidth reveals that the proposed receiver is very tolerant to transmitter phase noise. BER degrades moderately (from $4\times10^{-3}$ to $9\times10^{-3}$) when linewidth increases from 1 kHz to 200 kHz whilst typical coherent receivers demand the use of narrow linewidth lasers (< 30 kHz). Fig. 7 presents indicative scatterplots. Fig. 7a shows the constellation of the transmitted signal in the presence of a strong residual carrier which packages all QAM-16 symbols to the first quadrant. Fig. 7b shows the output constellation when filters without feedback are used. It becomes evident that it is not easy for the linear equalizer of the digital readout to classify the 16 symbols, which are characterized by strong nonlinearities in the constellation diagram. On the contrary, when recurrent filters are utilized (fig. 7c), the classification performance vastly improves using the same readout and the BER becomes equal to $5\times10^{-3}$.

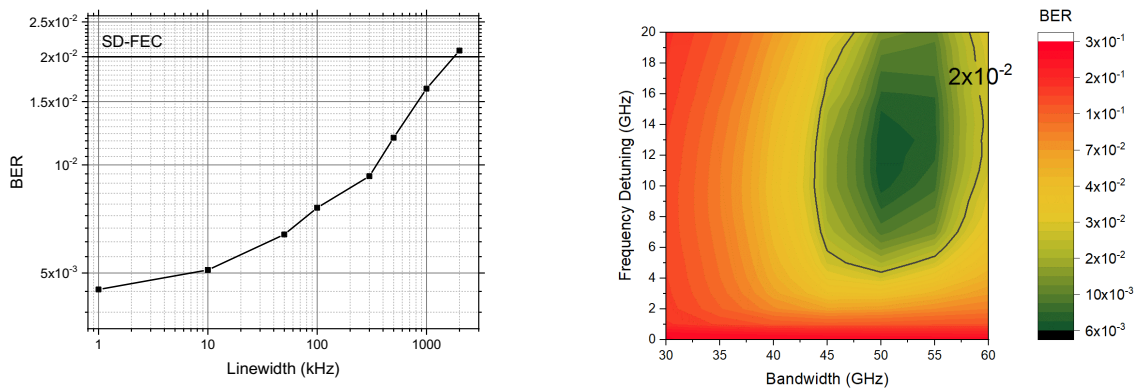

**Figure 6: The dependence of the coherent system's performance on transmitter linewidth and on the receiver parameters.** a) The BER performance is not severely affected by an exponential increase in the transmitter linewidth. This is attributed to the direct detection nature of the receiver. b) The contour plot of BER performance as a function of bandwidth, detuning for the 3 recurrent nodes with $L=0.5$, $T_d=9$ ps of the receiver. The first filter is centered on the carrier frequency while the other two are detuned symmetrically in the blueshifted and redshifted frequency components. The best results are achieved for 50 GHz bandwidth proving that spectral slicing relaxes the need for high bandwidth photodetectors and ADCs.

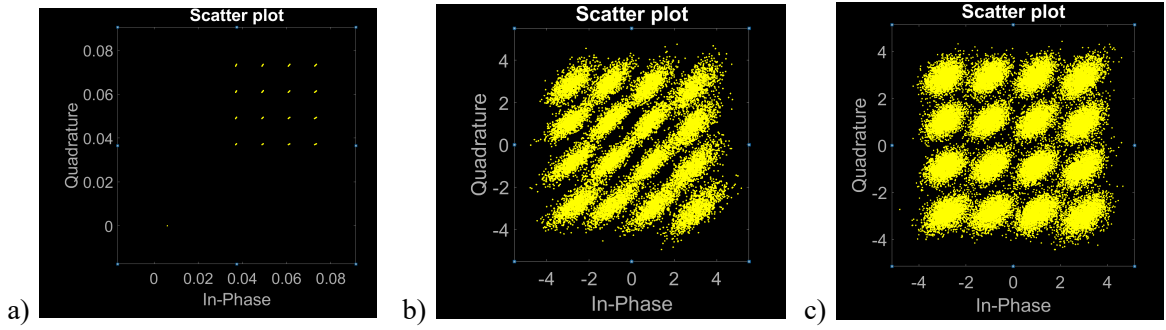

**Figure 7: Scatterplots of the transmitted and received QAM-16 signal** a) The transmitted QAM-16 signal with an increased residual carrier following the paradigm of self-coherent techniques b) The received signal after 20 km of O-band transmission with D=3 ps/nm/km and after being processed by 3 simple filters. The BER performance is in the order of $3\times10^{-2}$. c) The signal received and processed by 3 recurrent nodes with $BW$=50 GHz, Frequency detuning=15 GHz, $L$=0.5, $T_d$=9 ps, achieving BER performance near the HD-FEC limit ($5\times10^{-3}$).

### C. Memory Capacity

As already reported, the recurrent nature of the optical filter is anticipated to increase its fading memory, which is required in problems such as dispersion mitigation. In order to estimate the short-term memory of the recurrent node, we solve the memory capacity task. Memory capacity *MC* is derived from the memory function *m(i)* of eq. (3) [3], [4] and quantifies the amount of information from past symbols that a recurrent system can reproduce.

$$m(i) = \frac{(y(n-1)\ o_i(n))^2}{\sigma^2(y(n))\ \sigma^2(o_i(n))} \tag{3}$$

In (3), *y(n)* is a random input signal in the range [0, 1], $o_i(n)$ is system output at time *n* when the output weights are trained to reproduce the *i*-th past input signal *y(n-i)*. $\sigma^2$ is the variance. Then, the memory capacity *MC* is defined as the sum of *m(i)* as follows:

$$MC = \int_1^\infty m(i) \tag{4}$$

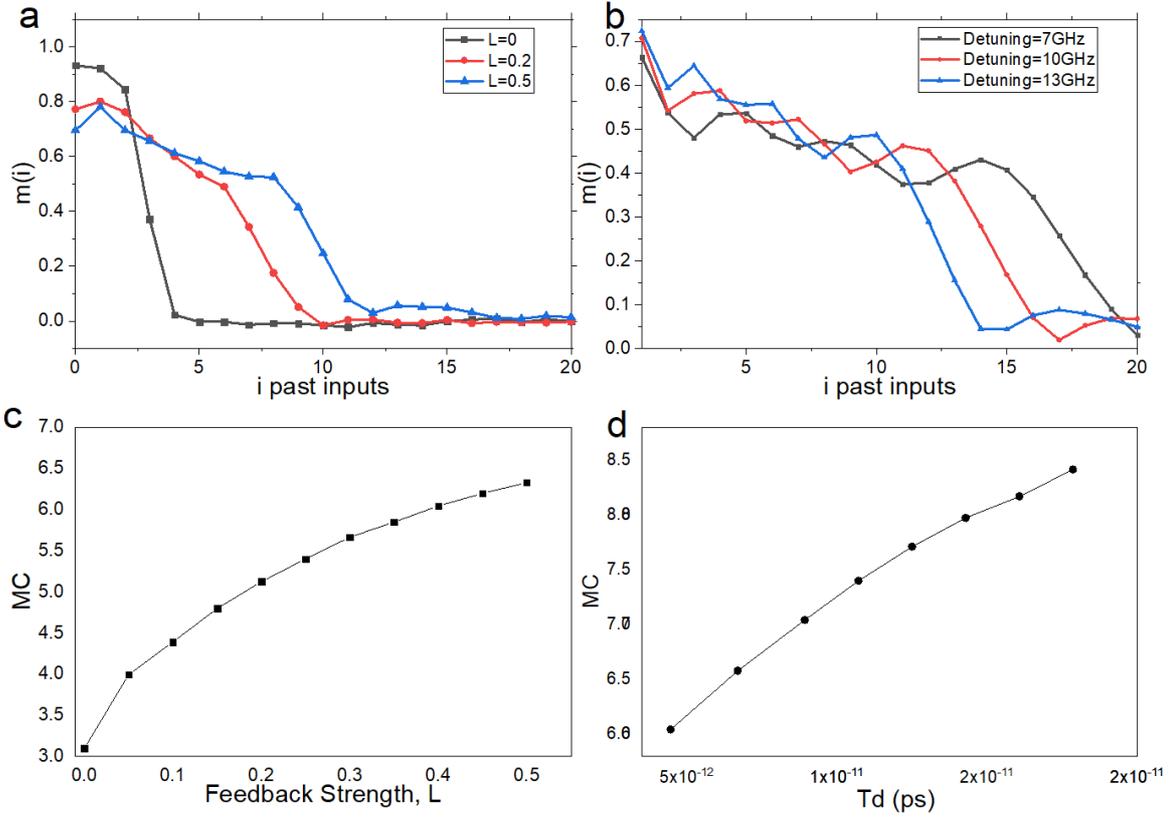

**Fig. 8. Memory capacity studies.** a) Memory function $m(i)$ in respect of different feedback strength values $L$ and b) of different frequency detuning values between the filters. c) The memory capacity of a system with four parallel recurrent nodes as a function of feedback strength $L$, and d) as a function of Time Delay of the feedback loop $T_d$. Tuning the two feedback parameters the memory ranges from 3 to 8.5 while the memory functions can spread even across 20 symbols in the past.

We calculate memory capacity for ROSS-NN considering $N_B=4$ and $N_F=1$, that is four recurrent nodes that slice input signal in four bands. We investigate the effect of feedback strength $L$, along with other important parameters such as filter bandwidth and frequency detuning. In fig. 8a we observe that $m$ increases when feedback strength is increased from zero to 50%. The fact that even without feedback a good correlation is achieved for one or two symbols is attributed to the intrinsic memory of the filter. Another important result arises when the frequency detuning between the nodes is tuned (fig. 8b). We consider four filters with detuning between the $i$-th and the $j$-th filter defined as $\Delta f_{ij}=(j-i)\Delta f$ with $\Delta f$ taking three values (7 GHz, 10 GHz, 13 GHz). All filters have 40 GHz 3dB bandwidth and the first filter is tuned to the central frequency of the incoming random signal. As $\Delta f$ scales from 7 to 13 GHz, we observe that fading memory extends to less symbols in the past. Therefore frequency detuning between different nodes also tunes the fading memory of the system. Fig. 8c calculates MC as a function of feedback strength. It is obvious that as feedback increases, the system is capable of holding information of more symbols from the past. The same trend is observed with the increase of feedback delay time (fig. 8d). Memory capacity scales as a square root function for the specific values of feedback delay time and frequency detuning. This square root behavior is observed in similar works [5].

.